# Unidirectional scattering induced by the toroidal dipole moment in the system of plasmonic nanoparticles

Lixin Ge, Liang Liu, Shiwei Dai, Jiwang Chai, Qianju Song, Hong Xiang, and Dezhuan Han*

*Department of Applied Physics, College of Physics, Chongqing University, Chongqing 400044, China*

dzhan@cqu.edu.cn;

**Abstract:** Unidirectional backward and forward scattering of electromagnetic waves by nanoparticles are usually interpreted as the interference of conventional multipole moments (i.e., electric and magnetic dipole, electric quadrupole, etc.). The role of toroidal dipole moments in unidirectional scattering is generally overlooked. In this work, we investigate the unidirectional scattering for the system of three plasmonic nanospheres. It is found that the unidirectional backward scattering is caused by the interference between the toroidal dipole moment and other conventional multipole moments. Tunable primary backward and forward scattering can be achieved under some specific configurations. Our results can find applications in the design of nanoantennas.

# 1. Introduction

Manipulating the directionality of the scattered light is of significance to many fields such as, nanoantennas [1], photovoltaic devices [2] and sensors [3], etc. The directionality of the scattered light is dependent on the excitation of the multipole moments of the scatterers. As is well known, for the Rayleigh scattering of a dielectric particle, the scattering pattern is symmetric for the forward scattering (FS) and backward scattering (BS) since only the electric dipole is excited [4]. However, by considering the magnetic dipole or other higher order moments involved in the scattering process, the unidirectional scattering can be realized, based on the destructive or constructive interference of multipole moments [5-12]. For a single particle made by magnetic material [5], high-index semi-conducting material [6-8], topological insulators [9] and spoof localized surface plasmon resonators [10], unidirectional scattering can be realized by utilizing the interference of electric dipole and magnetic dipole moments. The unidirectional scattering has also been demonstrated in core-shell particles [11-12] with quadrupole moments taken into account.

The unidirectional scattering of light by nanoparticles is usually interpreted as interference of conventional multipole moments (i.e., electric and magnetic dipole, electric quadrupole, etc.). However, there exists another kind of unusual multipole moments, called toroidal moments, which serve as the third family of multipole moments in addition to conventional electric and magnetic multipole moments [13]. The role of toroidal dipole moment in the unidirectional scattering is generally overlooked. The reason can be attributed to the fact that the toroidal moment is a high-order term and the corresponding radiation is relatively small for scatterers in the sub-wavelength scale. Recently, it is shown that the radiation of the toroidal dipole moment can be greatly enhanced in the structures of split-ring resonators [13-14], high index meta-materials [15] and other artificial structures [16] by mimicking a closed loop of magnetic dipoles. Resonant toroidal dipole mode can also be supported by an assembly of plasmonic nanoparticles as long as the electric dipole resonance exists for these particles [17]. Although the radiation pattern from the toroidal dipole is indistinguishable from that of the conventional electric dipole for an observer in the far field, the characteristics such as the resonance frequency and quality factor can be much different from those of the conventional electric dipole. Many interesting optical responses induced by the toroidal dipole moment are revealed, including, non-trivial optical transparency [18], anapole [19], lasing [20], all-optical Hall effect [21] and nanoantennas [22], etc.

In this letter, we show that toroidal dipole moment can indeed play an important role in the unidirectional scattering of an assembly of nanoparticles. The unidirectional backward scattering can be interpreted by the interference between the toroidal dipole moment and other conventional multipole moments.

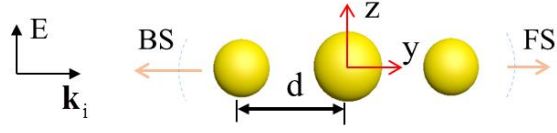

Fig. 1. Schematic view of the system under study. The incident wave is polarized along the z axis. The radii of three plasmonic nanospheres are $R_1$, $R_2$ and $R_3$, respectively, from left to right. The BS (backward scattering) and FS (forward scattering) are along -$k_i$ and $k_i$, respectively.

## 2. Multipole expansion method

Here, we consider the scattering properties of a system composed of three plasmonic nanospheres, which has been proved to be a platform for toroidal moments in a wide range of frequencies [17]. The system under study is shown in Fig. 1. Three plasmonic nanospheres are aligned along the y-axis. The center-to-center distance between the neighboring particles is $d$. The radius of the central sphere is $R_2$, different from the radii of the side ones ($R_1=R_3$). The background is the free space in our study. The permittivity for the plasmonic nanospheres is described by the Drude model with $\varepsilon(\omega) = 1 - \omega_p^2/(\omega^2 + i\gamma\omega)$, where $\omega_p$ is the plasma frequency and $\gamma$ is the electron scattering rate. We consider an incident plane wave propagating along the positive y axis with $\mathbf{E}^{ext} = \hat{e}_z E_0 e^{ik_i y}$, where $E_0$ is the amplitude, $k_i=\omega/c$ is the wave vector, $c$ is the speed of light in vacuum. The factor of time harmonics $e^{-i\omega t}$ is omitted here. The nanospheres can be treated as point dipoles since the radii of the nano-spheres we considered are much smaller than the incident wavelength. The induced dipole moment $\mathbf{p}_i$ ($i$=1, 2, 3) located at $\mathbf{r}_i$ can be determined by the following coupled dipole equation [23]:

$$\mathbf{p}_i = \alpha_i(\omega)[\sum_{j \neq i} \ddot{\mathbf{g}}(\mathbf{r}_i - \mathbf{r}_j)\mathbf{p}_j + \mathbf{E}_i^{ext}] \qquad (1)$$

where $\alpha_i(\omega) = (3i/2k_0^3) a_1(\omega)$ is the polarizability of $i$-th nanosphere, $a_1(\omega)$ is the electric dipolar term of the Mie coefficients [4], $\mathbf{E}_i^{ext}$ represents the external electric field on the $i$-th nanosphere, $\ddot{\mathbf{g}}(\mathbf{r}_i - \mathbf{r}_j)$ is the dyadic Green function. The dipole moment for each nanoparticle can be obtained by solving the above equation. As a result, the total multipole moments for the tri-particle assembly, such as the electric dipole $\mathbf{P}$, magnetic dipole $\mathbf{M}$, toroidal dipole $\mathbf{T}$, electric quadrupole $Q_e$, and the magnetic quadrupole $Q_m$ can also be obtained through the above-calculated point dipoles $\mathbf{p}_i$ as follows [17]:

electric dipole moment: $\mathbf{P} = \sum_i \mathbf{p}_i$ ,

magnetic dipole moment: $\mathbf{M} = \dfrac{-ik_0}{2} \sum_i \mathbf{r}_i \times \mathbf{p}_i$,

toroidal dipole moment: $\mathbf{T} = \dfrac{ik_0}{10} \sum_i \left[ 2r_i^2 \mathbf{p}_i - (\mathbf{r}_i \cdot \mathbf{p}_i) \mathbf{r}_i \right]$, electric quadrupole moment: $\ddot{Q}^{(e)}_{\alpha\beta} = \dfrac{1}{2} \sum_i \left[ r_{i,\alpha} p_{i,\beta} + r_{i,\beta} p_{i,\alpha} - \dfrac{2}{3} (\mathbf{r}_i \cdot \mathbf{p}_i) \delta_{\alpha\beta} \right]$, and magnetic quadrupole moment: $\ddot{Q}^{(m)}_{\alpha\beta} = -\dfrac{ik_0}{3} \sum_i \left[ (\mathbf{r}_i \times \mathbf{p}_i)_\alpha r_\beta + (\mathbf{r}_i \times \mathbf{p}_i)_\beta r_\alpha \right]$.

Other higher-order terms are negligible in our study. The radiation power from these different multipole moments can be easily achieved as follows [13, 21]:

$$I_p = \frac{2\omega^4}{3c^3} |\mathbf{P}|^2, \quad I_m = \frac{2\omega^4}{3c^3} |\mathbf{M}|^2, \quad I_T = \frac{2\omega^6}{3c^5} |\mathbf{T}|^2,$$

$$I_{Q_e} = \frac{\omega^6}{5c^5} \sum \left| \ddot{Q}^{(e)}_{\alpha\beta} \right|^2, \quad I_{Q_m} = \frac{\omega^6}{40c^5} \sum \left| \ddot{Q}^{(m)}_{\alpha\beta} \right|^2.$$

Meanwhile, the scattered electric field in the far field can be given by these multipole moments as follows [24-25]:

$$\mathbf{E}_s = R(r)[\mathbf{F}_p(\theta,\phi) + \mathbf{F}_T(\theta,\phi) + \mathbf{F}_m(\theta,\phi) + \mathbf{F}_{Q_e}(\theta,\phi) + \mathbf{F}_{Q_m}(\theta,\phi)] \quad (2)$$

where $R(r) = \omega^2 \mu_0 e^{ik_0 r}/4\pi r$ is the radial factor of the scattering electric fields, $r$ is the radial distance of the far-field detector, $\mu_0$ is the permeability of the vacuum, $\mathbf{F}(\theta, \phi)$ is the polar ($\theta$) and azimuthal ($\phi$) factor of the scattered electric field. The subscripts of $\mathbf{F}(\theta,\phi)$ stand for different multipole moment with $\mathbf{F}_p = \mathbf{n} \times (\mathbf{p} \times \mathbf{n})$, $\mathbf{F}_T = \mathbf{n} \times (ik_0 \mathbf{T} \times \mathbf{n})$, $\mathbf{F}_m = \mathbf{M} \times \mathbf{n}$, $\mathbf{F}_{Q_e} = ik_0 \mathbf{n} \times (\mathbf{n} \times \ddot{Q}^{(e)} \mathbf{n})/2$, and $\mathbf{F}_{Q_m} = ik_0 (\mathbf{n} \times \ddot{Q}^{(m)} \mathbf{n})/2$. Here $\mathbf{n} = \mathbf{r}/r$ is the unit vector in the radial direction. For the FS or BS, the unit vector $\mathbf{n}$ is parallel or opposite to the direction of incident wave vector $\mathbf{k}_i$, respectively. Noted that $\mathbf{F}_m(\theta, \phi)$ and $\mathbf{F}_{Q_e}(\theta, \phi)$ are odd functions of $\mathbf{n}$ and their signs will be flipped from FS to BS, whereas $\mathbf{F}_{p,T,Q_m}(\theta, \phi)$ are even function of $\mathbf{n}$ and the scattered electric fields generated by these three multipole moments are symmetric for BS and FS.

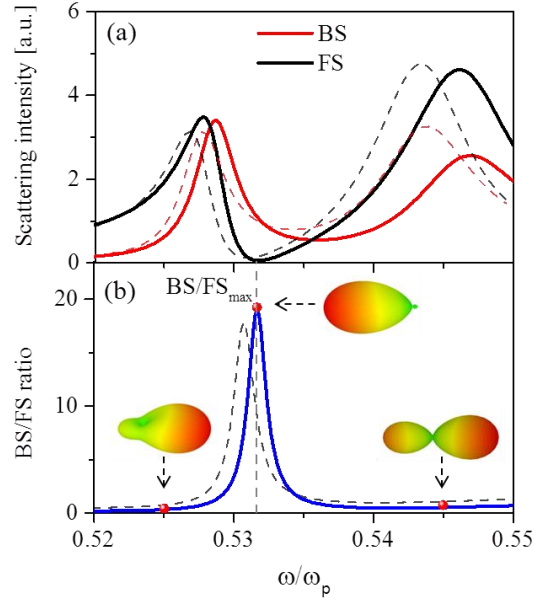

Fig. 2. (a) The scattering intensity of BS and FS as functions of frequency, calculated by the electric dipole approximation (solid lines) and COMSOL Multiphysics (dashed lines), respectively. (b) The ratio of BS to FS intensity. The blue solid line is calculated from the radiation of the multipole moments (**P**, **T**, **M**, **Q**$_e$ and **Q**$_m$) of the tri-particle assembly, while the dashed line is the simulated result. The insets show the angular distributions of scattering intensity in the far field for three different frequencies. Here we set $R_1 = R_3 = 20$ nm, $R_2 = 25$ nm, $d = 60$ nm.

## 3. Results and discussions

We can investigate the properties of scattered fields in the far field by summing up the contributions from all the multipoles in Eq. (2). The scattering intensity $I = E_s^* E_s$ versus the incident frequency for the FS and BS are shown in Fig. 2(a). Here, we set $R_1 = R_3 = 20$ nm, $R_2 = 25$ nm, $d = 60$ nm, $\omega_p = 6.18$ eV, and the nanospheres are assumed to be lossless with $\gamma = 0$. The spectra of scattering intensity for BS and FS are quite different. The maximal ratio of BS to FS intensity appears at the frequency of 0.532 $\omega_p$, at which the scattering intensity of FS has a dip. The ratio of BS intensity to FS intensity is about 20, which is much larger than the typical maximal ratio of BS to FS for a PEC sphere with 9:1 [26]. The peak of the maximal ratio of BS to FS is denoted by BS/FS$_{max}$ (dashed gray line). The results simulated by COMSOL Multiphysics are also shown for comparison. It is noted that there only exists a small deviation of frequency for BS/FS$_{max}$ between the analytical and numerical results. This small deviation comes from the finite-size effect of nanospheres. The angular distributions of the scattering intensity in the far field are shown in the insets of Fig. 2(b) for three different frequencies indicated by the red points.

To analyze the underlying mechanism of the unidirectional scattering, the radiation power from different channels of multipole moments are given in Fig. 3(a). The parameters are kept the same as those in Fig. 2. Two types of resonant modes in our system are efficiently excited. One mode with resonance frequency $\omega \sim 0.528\omega_p$ possessing a resonant peak of toroidal dipole moment has been demonstrated to be a special kind of symmetric mode with the side dipoles out of phase with the central one [17]. The other is a resonant magnetic mode (resonance frequency $\omega \sim 0.547\omega_p$). This is an asymmetric mode with the two side dipoles oscillating out of phase while the dipole moment of the central one vanishing. The frequency of BS/FS$_{max}$ locates between the resonances frequencies of these two modes, which is represented by the dashed gray line in Fig. 3(a). It is noted that the radiation power from the total electric dipole moment **P** almost becomes zero at the frequency of BS/FS$_{max}$. The corresponding near field distribution at the frequency of BS/FS$_{max}$ is shown in Fig. 3(b). The electric field has been enhanced remarkably near the surface of spheres. Meanwhile, a "toroidal-like" distribution of the magnetic field is observed clearly, manifesting the significant role that the toroidal moment plays in this unidirectional scattering. For the FS, the scattered electric field **E**$_s$ are decomposed into different channels of multipole moments as shown in Fig. 3(c) and 3(d). In Fig. 3(c), the amplitude of $\mathbf{F}_T+\mathbf{F}_{Q_m}$ are almost the same as that of $\mathbf{F}_m+\mathbf{F}_{Q_e}$ at the frequency of BS/FS$_{max}$ ($|\mathbf{F}_T+\mathbf{F}_{Q_m}|/|\mathbf{F}_m+\mathbf{F}_{Q_e}| \sim 0.9$). The corresponding scattered electric fields generated by these two pairs of multipole moments, namely, ($\mathbf{F}_T$, $\mathbf{F}_{Q_m}$) and ($\mathbf{F}_m$, $\mathbf{F}_{Q_e}$) are always in phase for the FS, however, the phase difference between these two pairs is nearly $\pi$ at the frequency of BS/FS$_{max}$ as shown in Fig. 3(d). Thus, they destructively interfere in the far field, leading to the significant suppression of FS. For the BS (not shown), the scattered electric fields generated by these two pairs, (**T**, $Q_m$) and (**M**, $Q_e$), constructively interfere in the backward direction since the phases of (**M**, $Q_e$) are reversed. Thus BS can be dramatically enhanced in the far field.

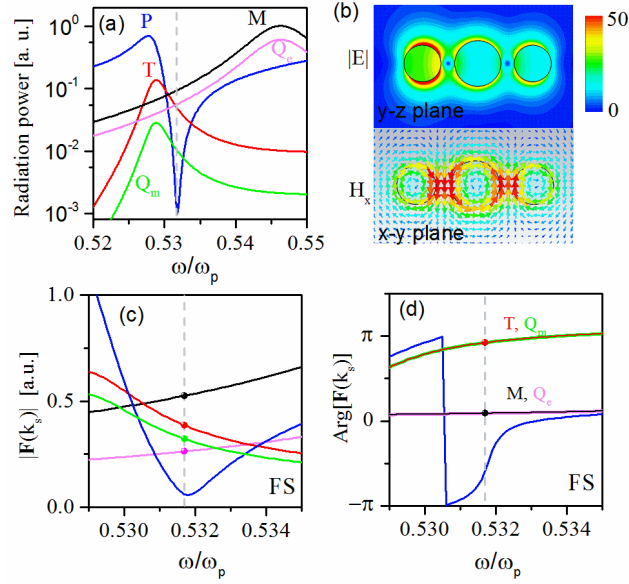

Fig. 3. (a) Radiation spectra of different multipole moments, including electric dipole (**P**), magnetic dipole (**M**), toroidal dipole (**T**), electric quadrupole($Q_e$) and magnetic quadrupole($Q_m$). (b) The spatial profile of the electric and magnetic field at the frequency of the maximum of BS/FS. The amplitude and phase of scattered electric fields corresponding to these multipole moments are shown in (c) and (d) in the forward direction, respectively. The colors in (c) and (d) represent the same multipole moments as those in (a). The dashed gray line indicates the frequency of the maximum of BS/FS (marked by BS/FS$_{max}$ in Fig. 2(b)). The geometric parameters are the same as those in Fig. 2.

Tunable directionality is highly desirable in the design of nanoantennas. In Fig. 4(a), the ratio of BS intensity to FS intensity versus the radius of central nanosphere $R_2$ and the incident frequency is given. The equal intensity with BS/FS=1 is marked by the black curves. The directionality of scattered waves of the tri-particle assembly can be different as we tune the radius of $R_2$. Under the condition $R_2>20$ nm, the frequency regime for the backward directionality becomes broaden as $R_2$ increases. Meanwhile, the maximal ratio of BS/FS becomes larger. For $R_2<20$ nm, however, the FS is dominant over the whole spectrum. Interestingly, the ratio of BS/FS can be extremely small (less than 0.02) as $R_2\sim21$ *nm*, represented by the black regime in Fig. 4(a), manifesting the significant suppression of BS. In Fig. 4(b), we plot the angular distribution of the scattering intensity in the far field (*y-z* plane) for the points A and B marked in Fig. 4(a). The angular distributions are completely different at these two frequencies. Strong suppression of BS is found at the point A, while the significant enhancement of BS is observed at the point B. The directionality here can also be interpreted by the interference of multipoles. The amplitude and phase of scattered electric fields radiated from each multipole moments are shown in Fig. 4(c) and 4(d). At the frequency of point A (orange dashed line), the amplitudes of the scattered electric fields contributed from **P** and **M**($Q_e$) are much larger than those from **T** and $Q_m$, therefore we can only consider **P** and **M**($Q_e$) at point A. The phase difference of scattered electric fields generated by **P** and **M**($Q_e$) are very small. Hence, they will constructively interfere in forward direction, yielding strong FS at point A. At point B (gray dashed line), the scattered electric field induced by **T** is the dominant one. However, the phase of scattered fields generated by **T** ($Q_m$) differ by almost $\pi$ compared to those of **P** and **M**($Q_e$) as shown in Fig. 4(d), leading to the destructive interference in the forward direction. As a result, the strongly enhanced BS can be observed.

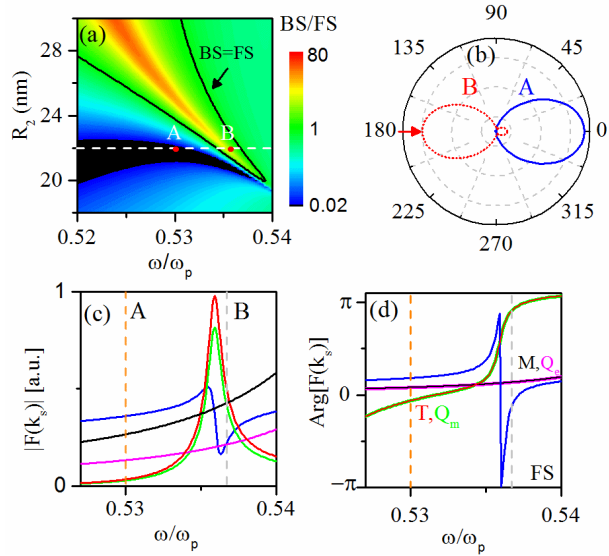

Fig. 4. (a) The ratio of BS intensity to FS intensity as a function of the radius $R_2$ of the central nanosphere and incident frequency. Other parameters are kept the same as those in Fig. 2. The black curves represent the equal intensity of BS and FS. The ratio of BS/FS is very low (<0.02) in the black colored area. (b) The angular distribution of the scattering intensity in the far field for the point A and B marked in (a), where the red arrow indicates the direction of incident waves. In (c) and (d), the amplitude and phase of scattered electric fields radiated by different multipole moments in the forward direction are shown for $R_2$=22 *nm* (corresponding to the dashed white line in (a)).

Finally, the intrinsic loss of plasmonic nanospheres are further considered. The ratio of BS/FS is shown in Fig. 5(a) with the loss taken into account. Compared with the lossless case with γ=0, the maximal ratio of BS/FS decreases as we increase the loss of plasmonic material. Nevertheless, the BS is still larger than the FS around the frequency of ω=0.528$ω_p$ even as

$\gamma=0.006\ \omega_p$. For the case of $\gamma=0.003\ \omega_p$, the radiation spectra of different multipole moments are shown in Fig. 5(b). The line shape is quite similar to those in Fig. 3(a). Again, we found that the radiation power from the toroidal dipole is larger than that of electric dipole at the frequency of $BS/FS_{max}$. In this case, the contribution of the toroidal dipole is essential in the unidirectional scattering.

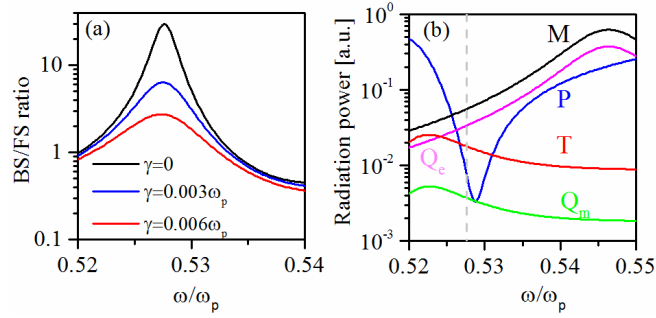

Fig.5. (a) The ratio of BS/FS with the intrinsic loss taken into account. (b) Radiation spectra for different multipole moments. The dashed gray line indicates the frequency of $BS/FS_{max}$. We choose $R_2=28$ *nm* and $\gamma=0.003\ \omega_p$, other parameters are kept the same as those in Fig. 2.

## 4. Conclusion

In summary, we study the unidirectional scattering for the system of three plasmonic nanoparticles. The asymmetric scattering between forward and backward direction is analyzed by the estimation of the radiation from all the multipole moments. We found that the unidirectional scattering is caused by the interference between the toroidal dipole moment and other conventional multipole moments. The maximal ratio of BS to FS can be interpreted by the fact that the scattered fields of **T**, $Q_m$ are out of phase with that of **M**, $Q_e$ (in some cases **P** is also involved) in the forward direction, while they are in phase in the backward direction. Our findings of unidirectional scattering may find applications in the design of nanoantennas, and also provide another way to investigate the electromagnetic properties of the elusive toroidal moment.

**Acknowledgements**

This work is supported the National Natural Science Foundation of China (Grant No. 11574037) and the Fundamental Research Funds for the Central Universities (Grant No. CQDXWL-2014-Z005 and 106112016CDJCR301205).